\documentclass[12pt,a4paper,dvips]{article}

\usepackage{amssymb,amsmath,graphicx,a4p}
\usepackage{cite,mcite}

\usepackage{ifthen}
\usepackage{l3_title}
\usepackage{Lep}
\usepackage{physics}

\journalname{Phys. Lett. B}
\date{February 08, 2000}
\preprint{2000-028}
\Lep{2}
\graphicspath{{./}}
\begin{document}
\begin{titlepage}
  
  \title{\boldmath Inclusive $\Sigma^+$ and $\Sigma^0$ Production in Hadronic Z Decays
    }

  \author{L3 Collaboration}
  
%
%
\begin{abstract}
  We report on measurements of the inclusive production rate of
  $\Sigma^+$ and $\Sigma^0$ baryons in hadronic Z decays collected with
  the L3 detector at LEP.
  The $\Sigma^+$ baryons are detected through the decay
  $\Sigma^+ \rightarrow {\rm p} \pi^0$, while the $\Sigma^0$ baryons are
  detected via the decay mode $\Sigma^0 \rightarrow \Lambda \gamma$.
  The average numbers of $\Sigma^+$ and $\Sigma^0$ per hadronic
  Z decay are measured to be:
  \begin{eqnarray*}
    \left< N_{\Sigma^+} \right> + \left< N_{\bar{\Sigma}^+} \right> & = & 0.114 \pm 0.011_{\mbox{\it \small stat}} \pm 0.009_{\mbox{\it \small syst}} \\
    \left< N_{\Sigma^0} \right> + \left< N_{\bar{\Sigma}^0} \right> & = & 0.095 \pm 0.015_{\mbox{\it \small stat}} \pm 0.013_{\mbox{\it \small syst}} \ \mbox{.}
  \end{eqnarray*}
  These rates are found to be higher than the predictions from
  Monte Carlo hadronization models
  and analytical parameterizations of strange baryon production.
\end{abstract}
  
\vspace*{1cm}
\centerline{Dedicated to the Memory of Howard D. Stone}
%
%
\submitted
\end{titlepage}
%
%
\section*{Introduction}
\label{sec:intro}
 
 Measurements of hadron production in e$^+$e$^-$ annihilation are important
 to understand the fragmentation process of quarks and gluons into hadrons.
 Perturbation theory cannot be applied to describe this
 process and the theoretical description is based on phenomenological
 models as implemented, for example, in the JETSET~\cite{JETSETuse},
 HERWIG~\cite{HERWIGuse} or ARIADNE~\cite{ARIADNE} Monte Carlo generators.

 Recently, two analytical models have been proposed to describe the hadron
 production rates in e$^+$e$^-$ annihilation~\cite{peiyj,becattini}.
 The model of Reference~\cite{peiyj}, referred to from here on as the
 ``string-based model,''
 is derived partially from string fragmentation
 and describes the production of light
 mesons and baryons using the simple formula:
 \begin{eqnarray}
   \left< N \right> = \frac{C(2J+1)}{C_B} \left(\gamma_s \right)^{N_s} e^{-\frac{E_{\rm bind}}{T}},
   \label{eqn:peiyj}
 \end{eqnarray}
 where $\left< N \right>$ is the number of primary hadrons of spin $J$,
 containing $N_s$ strange quarks, produced directly by string fragmentation.
 The quantity $E_{\rm bind}$ is the hadron binding energy:
 $E_{\rm bind} = M_{\rm{hadron}}- \sum_{i} m_{q_i}$,
 where $m_{q_i}$ are constituent quark masses, and $T$ is the effective
 temperature of hadronization in this model.  The overall normalization
 of the rate is controlled by $C$.  The evolution of $C$ with center-of-mass
 energy is predicted by perturbative QCD~\cite{QCDCM}.
 The parameter $\gamma_s$ is a universal suppression
 factor for strange quark production, while $C_B$ is a suppression factor
 of baryon production relative to meson production.
 This model has five free parameters: $C$ at some fixed energy
 scale, $C_B$, $\gamma_s$, $T$ and $\Delta m = m_s - m_u$, the mass difference
 between strange and up quarks.
 The model of Reference~\cite{becattini}, referred to from here on as the
 ``hadron gas model,'' is based on thermodynamical considerations and has
 three adjustable parameters: the total volume
 $V$ of the hadron gas, a strangeness suppression factor similar to
 $\gamma_s$ in Reference~\cite{peiyj}, and the hadronization temperature $T$. 
 In the hadron gas model, no additional parameters are needed to reproduce
 the relative production rate of baryons versus mesons.
 Both of these models give a good description of the full spectrum of
 light hadron production in addition to predicting the rates for strange
 baryons.  The string-based model also describes well the relative
 production rate of hadrons containing b and c quarks.
 In order to compare the experimental data to these two analytical
 models which predict primary hadron production, the observed rate must
 first be corrected for two effects: (i) hadrons containing
 quarks produced in the primary electroweak interaction, (ii)
 `feed-down' due to the decays of unstable hadrons produced in string
 fragmentation~\cite{peiyj} or, correspondingly, the phase change that
 produces the hadron gas~\cite{becattini}.  The feed-down correction is
 made using known hadron branching ratios~\cite{PDG1998}.  

 This paper presents measurements of $\Sigma^+$ and $\Sigma^0$ 
 strange baryons produced in hadronic Z decays at center-of-mass 
 energies in the range $\sqrt{s} = 91 \pm 2$~\GeV{}, from 71.1~pb$^{-1}$
 of data collected
 by the L3 detector during the LEP running periods of 1994 and 1995.
 The $\Sigma^+$ and $\Sigma^0$ baryons
 provide a good test of the hadron rate models because of their relatively
 low feed-down correction as compared to the lighter p and $\Lambda$ baryons.
 The use of the decay modes which have photons, 
 $\Sigma^+ \rightarrow {\rm p} \pi^0$ and
 $\Sigma^0 \rightarrow \Lambda \gamma$, enable the measurements to be
 made down to low baryon momenta due to the low-energy photon detection
 capability of the L3 detector.
%
%
\section*{The L3 Detector}

\begin{figure}[htb]
\begin{center}
    \includegraphics[width=11.0cm]{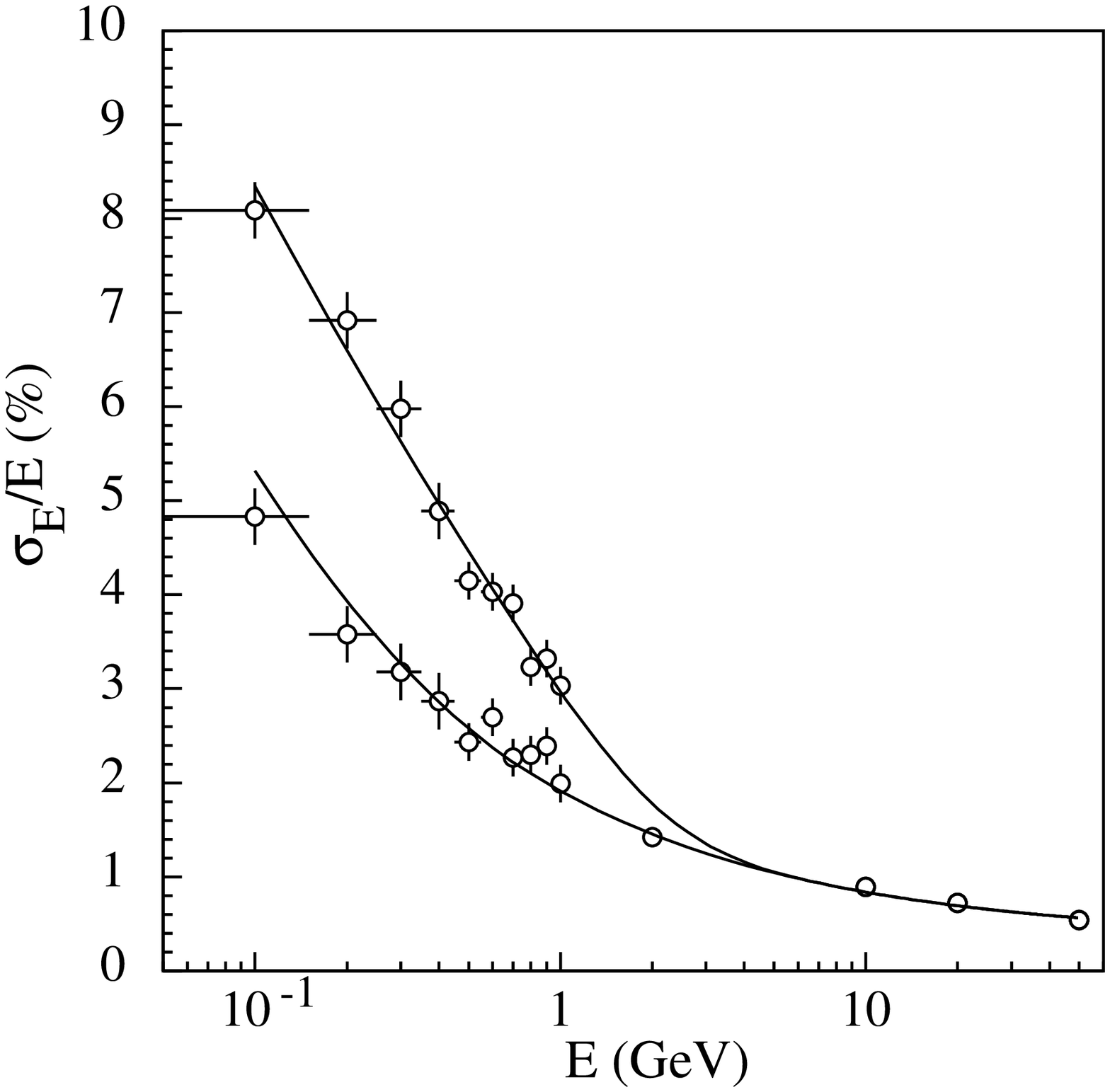}
\setlength{\unitlength}{1pt}
    \begin{picture}(300,0)(0,0)
      \put(250,260){\Large a)}
      \put(57,74){\large Gaussian}
      \put(57,57){\large }
      \put(75,90){\vector(1,1){20}}
      \put(148,202){\large Full Resolution}
      \put(144,201){\vector(-1,-1){20}}
      \put(250,-30){\Large b)}
      \put(112,-45){\large }
      \put(105,-62){\large Crystal Size}
      \put(95,-62.5){\vector(-1,-1){26}}
    \end{picture}
\setlength{\unitlength}{1pt}
    \includegraphics[width=11.0cm]{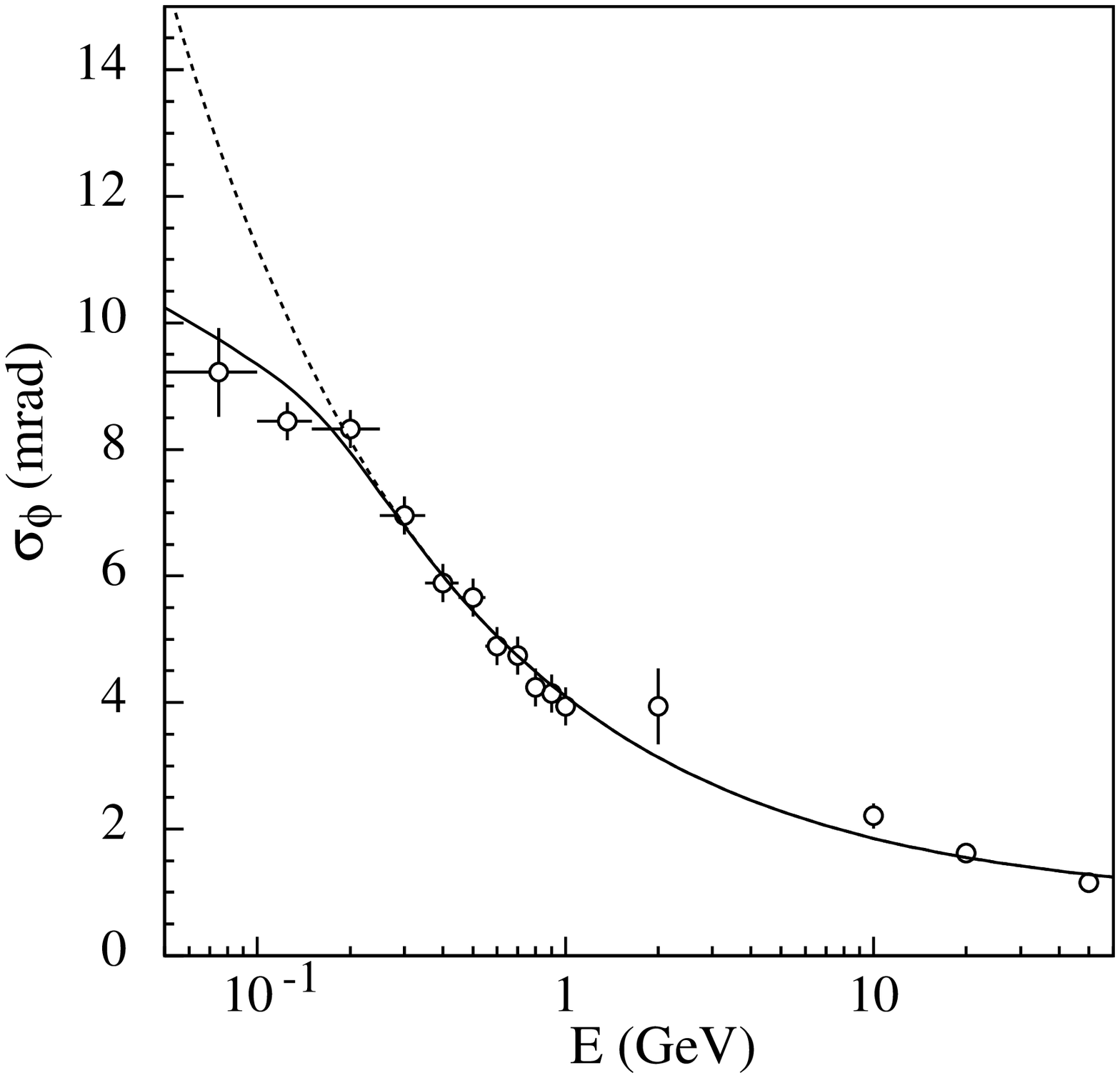}
\caption{Resolution of electromagnetic showers in energy and 
azimuthal angle $\phi$.  In plot a)
the two curves describe the energy resolution in the BGO barrel.
The full resolution curve accounts for the fraction of measured photons
which lack full shower containment.  These give a tailing distribution with
less resolution than the intrinsic Gaussian resolution curve.
In plot b) is the $\phi$-resolution function.  At low-energy, the angular
resolution is limited by the crystal size.  The dotted line shows the
extrapolated angular resolution when this effect is ignored.}
\label{fig:bgores}
\end{center}
\end{figure}

A detailed description of the L3 detector and its performance
is given in Reference~\cite{l3-00}.  This analysis makes use of
the electromagnetic calorimeter made of Bismuth Germanate~(BGO) crystals,
which accurately identifies and measures photons from an energy of
50~\MeV{} to 200~\GeV{}.
In order to benefit from the low-energy photon detection of the calorimeter,
its performance was studied and parameterized~\cite{pri97}.
The resulting resolution functions are shown in Figure~\ref{fig:bgores}.
In addition, corrections due to electronic noise and calibration accuracy
are taken into account.
The electronic noise terms are $\sqrt{N} \cdot \sigma_{\rm intr}$ with
$\sigma_{\rm intr} = 1$~\MeV{} and $N \cdot \sigma_{\rm corr}$
with $\sigma_{\rm corr} = 1.6$~\MeV{} 
where $\sigma_{\rm intr}$ is the mean intrinsic noise per channel and $\sigma_{\rm corr}$ is the mean correlated noise per channel, and 
$N$ is the number of crystals used to compute the shower energy.
The calibration accuracy term for
1994 and 1995
is $\sigma_{\rm cal} \cdot E$ with
$\sigma_{\rm cal} = 1.5\%$ as estimated from Bhabha events.
These terms are added in quadrature
to the resolution functions of Figure~\ref{fig:bgores}.
%
%
\section*{Hadronic Event Selection}

The inclusive production rate measurements are based
on a selection of hadronic Z decays which have a measured primary
vertex.  A precise measurement of the primary vertex in all three
dimensions is achieved in more than 99\% of all hadronic events.
The hadronic selection requires a
well-balanced, high particle-multiplicity event with the full
collision energy measured by the detector and relies on
the following criteria: 
{
\begin{itemize}
\item[1.] The total energy observed in the detector, including the momenta
of muons measured in the muon spectrometer, $E_{\rm vis}$, is restricted
to the range $0.6 \leq E_{\rm vis}/\sqrt{s} \leq 1.4$.
\item[2.] The energy imbalance along the beam direction, $E_{\|}$,
must satisfy $\left| E_{\|} \right| /E_{\rm vis} \leq 0.4$.
\item[3.] The transverse energy imbalance, $E_{\bot}$, with respect
to the beam direction, must satisfy
$E_{\bot}/E_{\rm vis} \leq 0.4$.
\item[4.] The number of energy clusters reconstructed
in the calorimeters is required to be larger than 12.
\end{itemize}
}

A data set of 1.6 million hadronic Z decays was selected for this
analysis.
%
%
\section*{Analysis Procedure}
\label{sec:fitproc}

The events passing the hadronic event selection are analyzed for the
inclusive production of $\Sigma^+$ and $\Sigma^0$ in the decay modes
${\rm p} \pi^0(\pi^0 \rightarrow \gamma \gamma)$ and 
$\Lambda (\rightarrow {\rm p} \pi^-) \gamma$, respectively.
The p$\pi^0$ and $\Lambda \gamma$ mass distributions are computed, and
a fitting procedure is applied to count the number of selected candidates
in the mass peak.  The anti-particles of the $\Sigma^+$ and $\Sigma^0$
are included in the corresponding distributions to increase the
statistical significance of the data samples.

Central to the selections of the $\Sigma^+$ and $\Sigma^0$ candidates
are the kinematical constraints describing their decay.
A probability is assigned to
each candidate based on the minimum $\chi^2$ which satisfies
the decay constraints.  The $\chi^2$ has
the form:
\begin{eqnarray}
\chi^2 & = &
           \sum_{i=1}^{N_{\rm tracks}}
           (\vec{t}_i - \vec{f}(\vec{v_s},\vec{q}_i))^T G^{-1}_i
           (\vec{t}_i - \vec{f}(\vec{v_s},\vec{q}_i))
         + (\vec{v} - \vec{v}_{\rm event})^T V^{-1}_{\rm event}
           (\vec{v} - \vec{v}_{\rm event}) \nonumber \\
   &  &    + \sum_{i=1}^{N_{\rm clusters}} 
           (\vec{u}_i - \vec{g}(\vec{v_s},\vec{q}_i))^T E^{-1}_i
           (\vec{u}_i - \vec{g}(\vec{v_s},\vec{q}_i))
\label{eqn:chi2}
\end{eqnarray}
where the number of charged-particle tracks and electromagnetic
clusters used to reconstruct the decay are respectively $N_{\rm tracks}$
and $N_{\rm clusters}$;
$\vec{t}_i$ is the vector of measured parameters for the $i^{\rm th}$
track and $G_i$ the corresponding covariance matrix; $f(\vec{v_s},\vec{q}_i)$
are the fit parameters assuming the track originated from the
secondary vertex $\vec{v_s}$ with momentum $\vec{q}_i$;
$\vec{v}_{\rm event}$ is the event-vertex with covariance matrix $V$;
$\vec{v}$ is the fit to the production vertex;
$\vec{u}_i$ is the vector of measured parameters for the $i^{\rm th}$
electromagnetic cluster with covariance matrix $E_i$; and
$\vec{g}(\vec{v_s},\vec{q}_i)$ are the fit parameters assuming that a photon
originated from the secondary vertex $\vec{v_s}$ with momentum $\vec{q}_i$.
The details of the constrained fitting technique are found in
Reference~\cite{pri97}.
In addition to the constraint probability cuts, other selection criteria
are also applied to reduce combinatorial background.

The fitting procedure of the invariant mass distributions
consists of several steps.
The JETSET Monte Carlo simulation is used to predict the shape of the
p$\pi^0$ and $\Lambda \gamma$ mass distributions separately for
the background and the $\Sigma^+$ and $\Sigma^0$ signals.
The Monte Carlo signal and background distributions are binned into
histograms which are then smoothed with a spline fit.  The spline fit
reduces background fluctuations in the signal region by interpolating
between the sidebands of the signal.
The use of binned histograms with bin widths comparable to the expected
mass resolution of the $\Sigma^+$ and $\Sigma^0$ signals is found to
be optimal for exploiting the histogram shape differences
between signal and background.
A maximum likelihood fit is performed for both the signal and the background
normalizations, simultaneously.  The likelihood is constructed by computing
the Poisson probability for each bin and then making the product over all
bins in the histogram, as follows,
\begin{eqnarray}
L(s,b) = e^{-(s+b)} \prod^{N_{\rm bins}}_{i=1} \frac{(f_i + g_i)^{d_i}}
{\Gamma(d_i+1)}
\end{eqnarray}
where $s$ and $b$ are respectively the total number of signal and background
events, $d_i$ is the number of observed data events in a given bin and $f_i$
and $g_i$ are the number of expected events in a given bin for signal and
background, respectively.

The production rates of the $\Sigma^+$ and $\Sigma^0$ are determined
by a kinematic extrapolation which assumes the JETSET Monte Carlo production
distributions.
%
%
\section*{$\Sigma^+$ Selection}

\begin{figure}[htb]
\begin{center}
\includegraphics*[width=8cm]{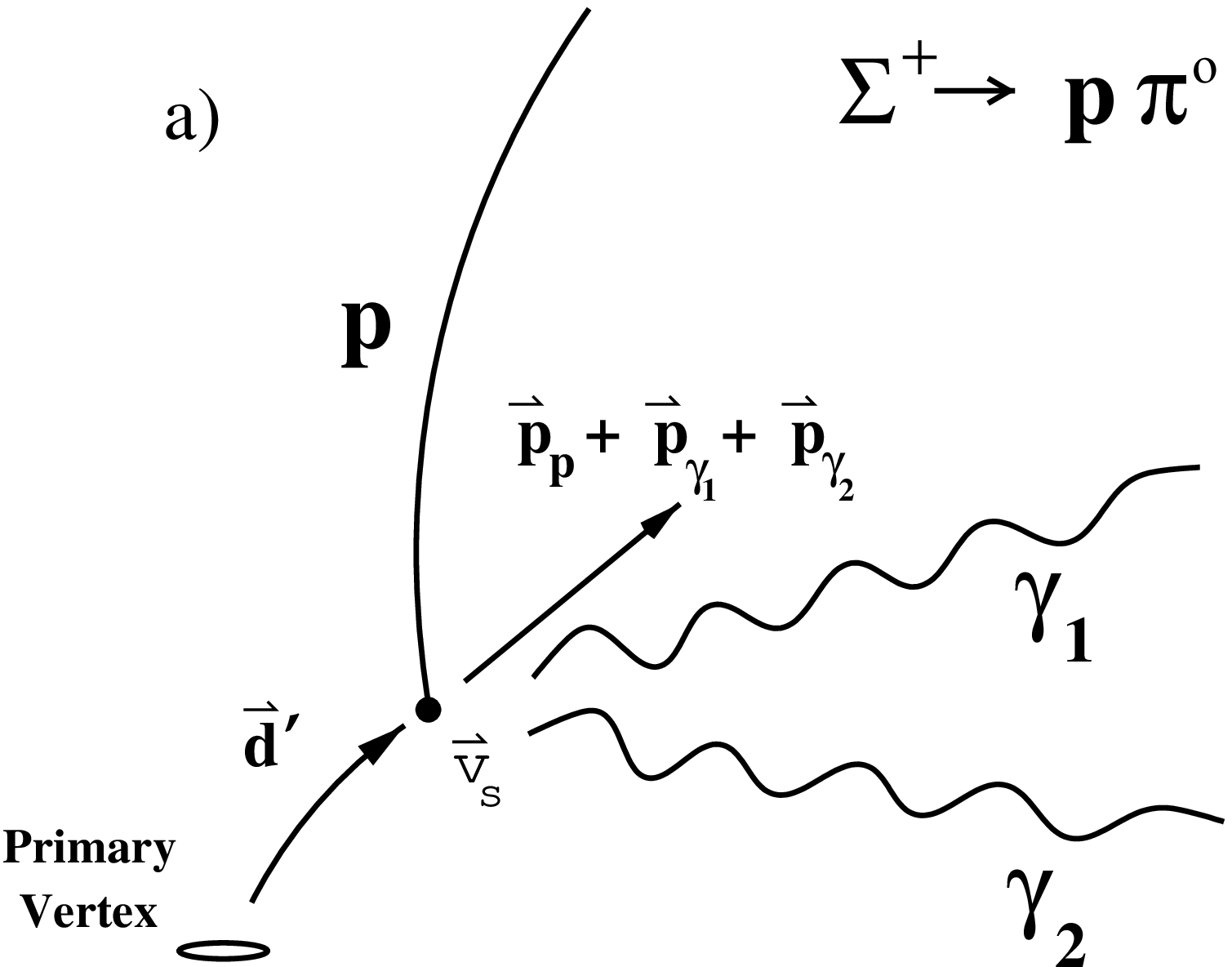}
\vspace*{5mm}
\end{center}

\vspace{15pt}
\begin{center}
\includegraphics*[width=8cm]{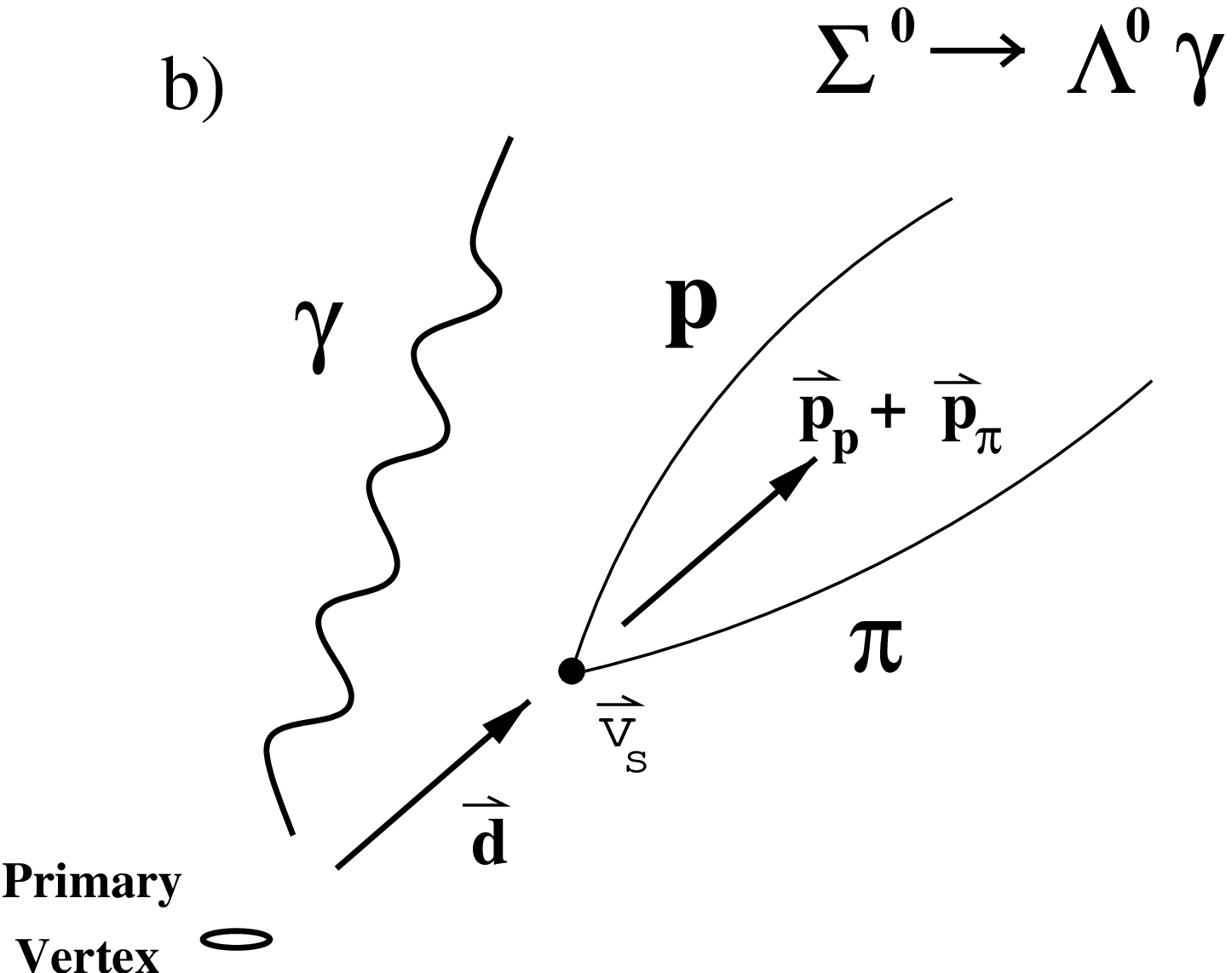}
\vspace*{5mm}
\caption{The decay of a $\Sigma^+$ is shown in plot a).
The position of the decay vertex, $\vec{v}_s$, and the path of the
$\Sigma^+$ from the primary vertex to the decay vertex are indicated.
The vector $ \rm \vec{\bf d'}$ is the direction of flight of the $\Sigma^+$
at the time of decay.
The decay of a $\Sigma^0$ is shown in plot b).
The point $\vec{v}_s$ indicates the position of the decay vertex
of the $\Lambda$, and the vector
$ \rm \vec{\bf d}$ is the direction of flight of the $\Lambda$.}
\label{fig:kinem}
\end{center}
\end{figure}

The signature of a
$\Sigma^+ \rightarrow {\rm p} \pi^0(\pi^0 \rightarrow \gamma \gamma)$ decay
is a single
charged track and two photons measured in the calorimeter as shown in 
Figure~\ref{fig:kinem}a.
The kinematical information of the decay configuration,
in addition to the assumption that the $\Sigma^+$
is produced at the primary vertex, can be expressed as follows:
\begin{eqnarray}
     \vec{\rm \bf d'} \times \left[ \vec{\rm \bf p}_{\rm p} 
     + \vec{\rm \bf p}_{\gamma_1}
     + \vec{\rm \bf p}_{\gamma_2} \right] = 0
\label{eqn:spcon1}
\end{eqnarray}
\begin{eqnarray}
     m_{\gamma \gamma} = m_{\pi^0}
\label{eqn:spcon2}
\end{eqnarray}
where the vector $\vec{\rm \bf d'}$ is the direction of flight of the
$\Sigma^+$ at the time of decay; the vectors $\vec{\rm \bf p}_{\rm p}$,
$\vec{\rm \bf p}_{\gamma_1}$ and $\vec{\rm \bf p}_{\gamma_2}$ are 
respectively the proton momenta and the momenta of the two photons;
and $m_{\gamma \gamma}$ and $m_{\pi^0}$ are the calculated mass of the
diphoton system and the $\pi^0$ mass, respectively.
A probability, $\mathcal{P}_{\mbox{\scriptsize svtx}}$, is assigned to
each $\Sigma^+$ candidate based on the minimum $\chi^2$ which satisfies
constraints~(\ref{eqn:spcon1}) and (\ref{eqn:spcon2}).  The $\chi^2$
is obtained from equation~(\ref{eqn:chi2}).
The computation of the constraint
probability requires the accurate description of the photon resolution
in energy and angle in the calorimeter.

The selection of $\Sigma^+$ candidates consists of the following cuts:
{
\begin{itemize}
\item[1.]
 The constraint probability, $\rm {\mathcal{P}_{\mbox{\scriptsize svtx}}}$,
 must be larger than 0.2.
\item[2.]
The energies of the two photons must be greater than $55$~\MeV{} and their polar
angle must be in the BGO barrel acceptance ($\rm \left| \cos \theta_{\gamma} \right|< 0.74$).
\item[3.]
The $\pi^0$ boost, $\gamma_{\pi^0} = E_{\pi^0}/m_{\pi^0}$, is required to be
larger than 2.0.
\item[4.]
The transverse momentum of the proton
must exceed 800~\MeV{}.
\item[5.]
The transverse decay radius, $r_{\bot}$, of the $\Sigma^+$
is required to be
in the range $5~\mbox{mm} < r_{\bot} < 70$~mm ensuring the decay to be inside
the inner radius of the Silicon Microvertex Detector (SMD).
\item[6.]
Charged tracks which are identified as coming from $\Lambda$ or K$^0_{\rm s}$
decay are rejected as proton candidates from $\Sigma^+$ decay.
\end{itemize}
}

\begin{figure}[htb]
\begin{center}
\includegraphics*[width=15.0cm,bb=10 25 530 525]{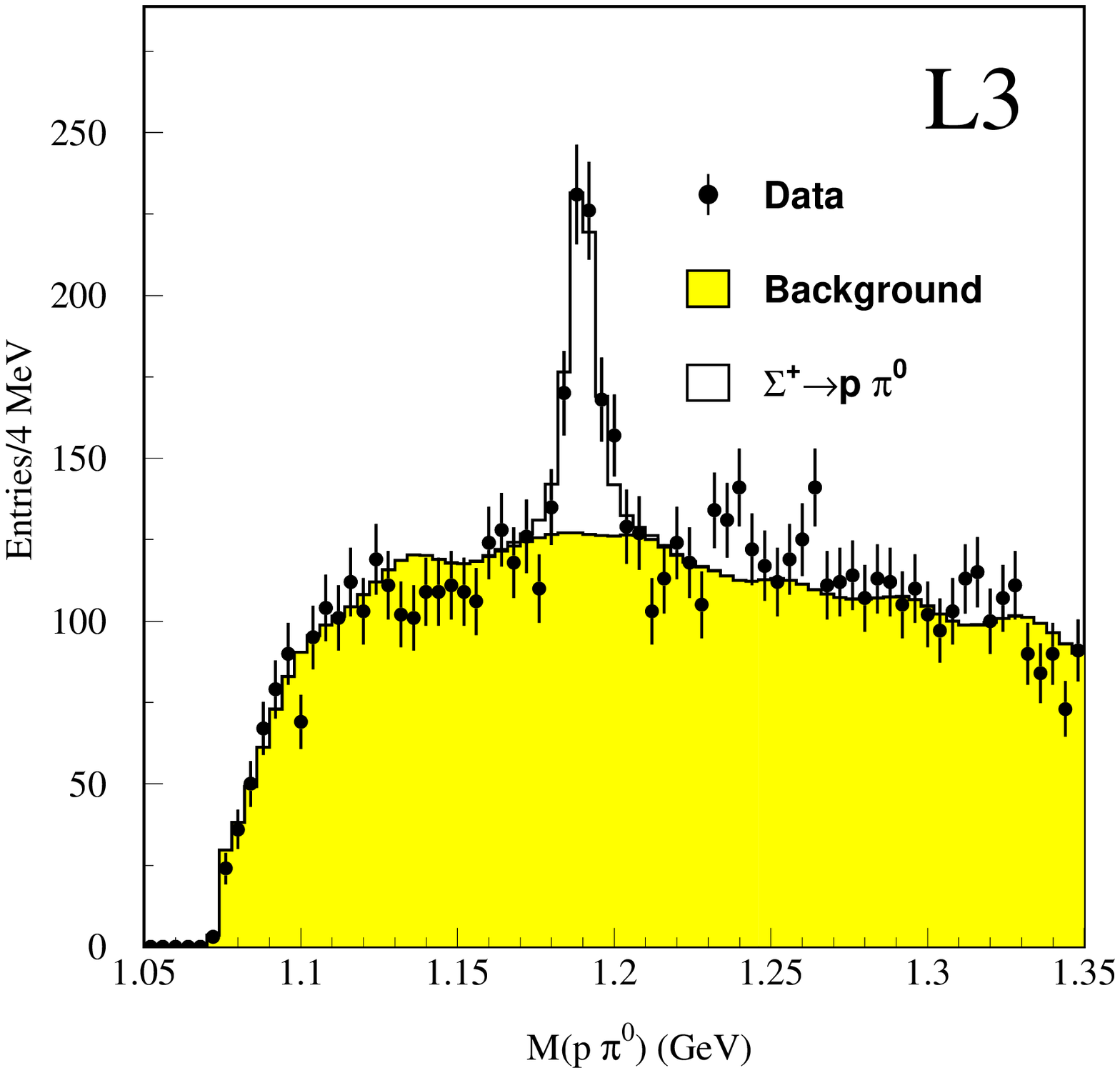}
\vspace*{-10pt}
\caption{The mass distribution of $\Sigma^+$ candidates.
The shapes of the signal and of the background are obtained from
Monte Carlo simulation as described in the text.}
\label{fig:newsp}
\end{center}
\vspace{-15pt}
\end{figure}

Figure~\ref{fig:newsp} shows the mass distribution of the selected
${\rm p} \pi^0$ candidates.
The number of selected $\Sigma^+ \rightarrow {\rm p} \pi^0$ decays
determined by the fit is found to be 342 $\pm$ 33.
%
%
\section*{$\Sigma^0$ Selection}

The $\Sigma^0$ is measured in the decay mode
$\Sigma^0 \rightarrow \Lambda \gamma$.  This mode accounts for nearly
100\% of the $\Sigma^0$ branching fractions.  
Identification of the $\Lambda$ is done using the
${\rm p} \pi^-$ decay mode.  The proton is assumed to be the track
with the highest momentum of the two tracks forming the $\Lambda$
candidate.
The kinematic constraints of the
$\Sigma^0 \rightarrow \Lambda (\rightarrow {\rm p} \pi^-) \gamma$ decay,
shown in Figure~\ref{fig:kinem}b, are summarized by the following equations:
\begin{eqnarray}
       \vec{\rm \bf d} \times \left[ \vec{\rm \bf p}_{\rm p} 
     + \vec{\rm \bf p}_{\pi^-} \right] = 0
\label{eqn:szcon1}
\end{eqnarray}
\begin{eqnarray}
m_{p \pi^-} = m_{\Lambda}
\label{eqn:szcon2}
\end{eqnarray}
where the vector $\rm \vec{\bf d}$ points from the primary vertex to the
$\Lambda$ decay point; the vectors $\vec{\rm \bf p}_{\rm p}$ and
$\vec{\rm \bf p}_{\pi^-}$ are respectively the proton and pion momenta;
and $m_{p \pi^-}$ and $m_{\Lambda}$ are the calculated mass of the
proton-pion system and the $\Lambda$ mass, respectively.
A probability, $\mathcal{P}_{\mbox{\scriptsize lvtx}}$, is assigned to
each $\Lambda$ candidate based on the minimum $\chi^2$
(equation~\ref{eqn:chi2}) which satisfies
constraints~(\ref{eqn:szcon1}) and (\ref{eqn:szcon2}).
A large combinatorial background to the $\Lambda$ signal is present
in the limit that
\begin{eqnarray}
       \vec{v}_s = \vec{v}_{\rm event} 
\label{eqn:szcon3}
\end{eqnarray}
where $\vec{v}_s$ is the $\Lambda$ decay vertex and $\vec{v}_{\rm event}$
is the primary event vertex.
A constraint probability,
$\rm \mathcal{P}_{\mbox{\scriptsize pvtx}}$,
is also computed for $\Lambda$ candidates to satisfy
equation~(\ref{eqn:szcon3}).  The $\Lambda$
signal has less background when $\rm \mathcal{P}_{\mbox{\scriptsize pvtx}}$
is small.

\begin{figure}[htb]
\begin{center}
\includegraphics*[width=15.0cm,bb=10 25 530 525]{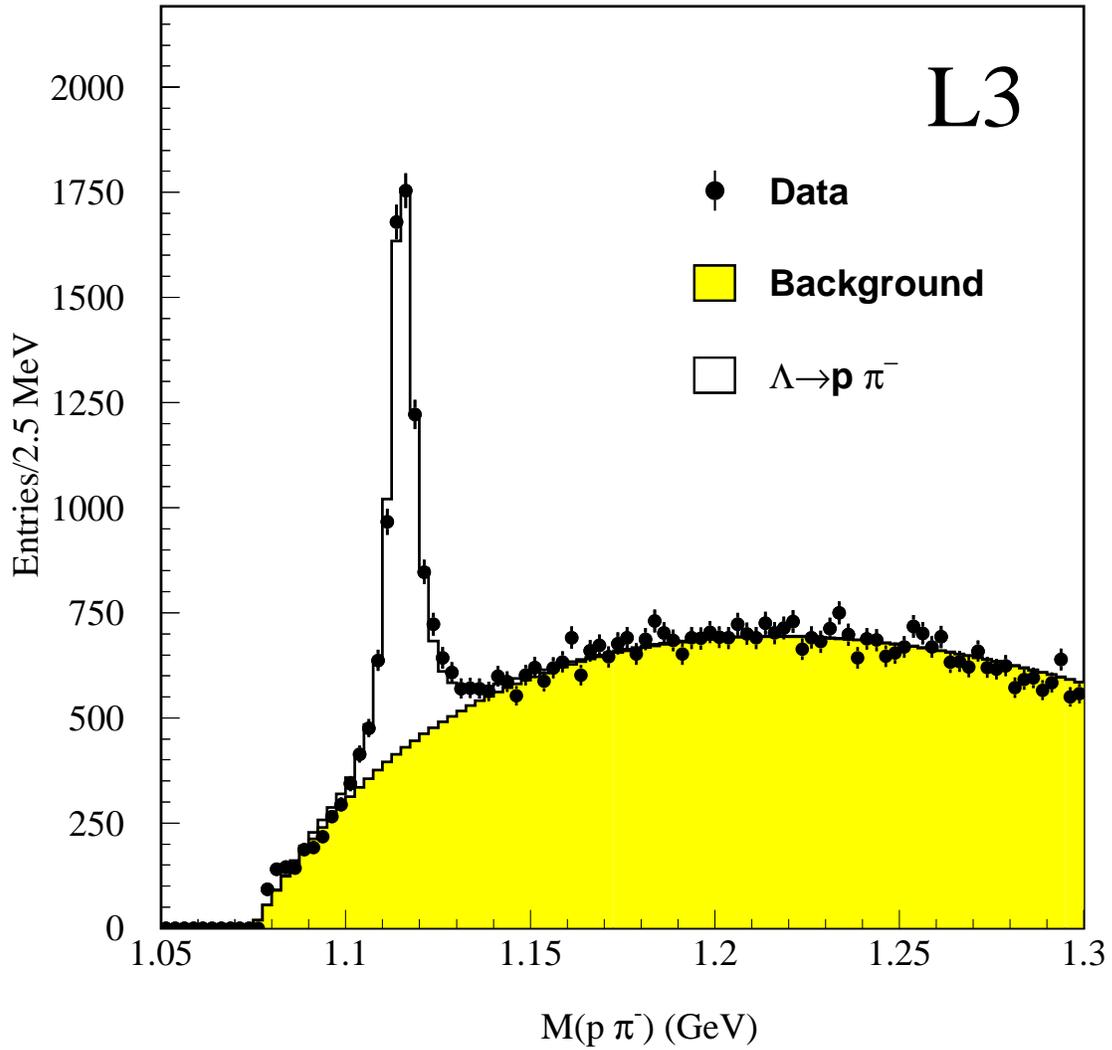}
\vspace*{-10pt}
\caption{The p$\pi^{-}$ mass distribution showing the $\Lambda$
candidates used to measure the $\Sigma^0$.}
\label{fig:lambda}
\end{center}
\vspace{-15pt}
\end{figure}

The $\Lambda$ selection consists of the following cuts:
{
\begin{itemize}
\item[1.]  The transverse momenta of the proton
and pion coming from the $\Lambda$ decay must both exceed 200~\MeV{}.
\item[2.]
 The constraint probability, $\rm {\mathcal{P}_{\mbox{\scriptsize lvtx}}}$,
 must be larger than 0.01, and
 the constraint probability, $\rm {\mathcal{P}_{\mbox{\scriptsize pvtx}}}$,
 must be smaller than $10^{-12}$.
\item[3.]
The transverse decay radius, $r_{\bot}$, of the $\Lambda$
is limited to the range $r_{\bot} < 70$~mm
ensuring the decay to be inside
the inner radius of the SMD.
\end{itemize}
}
After applying the $\Lambda$ selection cuts with the exception
of constraint~(\ref{eqn:szcon2}),
the reconstructed ${\rm p} \pi^-$ mass spectrum is plotted in
Figure~\ref{fig:lambda}
where the $\Lambda$ is clearly visible.
To reconstruct the $\Sigma^0$ decay, an electromagnetic cluster in the
BGO detector is combined with a selected $\Lambda$.

The following additional criteria is applied to select $\Sigma^0$ candidates:
{
\begin{itemize}
\item[1.]  The energy of the photon, $E_{\gamma}$, is restricted
to the range $ 55~\mbox{\MeV{}} < E_{\gamma} < 1$~\GeV{} and the photon polar
angle must be
within the BGO barrel acceptance ($\rm \left| \cos \theta_{\gamma} \right|< 0.74$).
\item[2.]  The angle, $\rm \theta^\ast_{\Lambda}$, of the
$\Lambda$ with respect to the $\Sigma^0$ momentum
in the $\Sigma^0$ rest frame
is required to satisfy 
$\rm \cos \theta^\ast_{\Lambda} < 0.2$.
\end{itemize}
}

\begin{figure}[htb]
\begin{center}
\includegraphics*[width=15cm,bb=10 25 530 525]{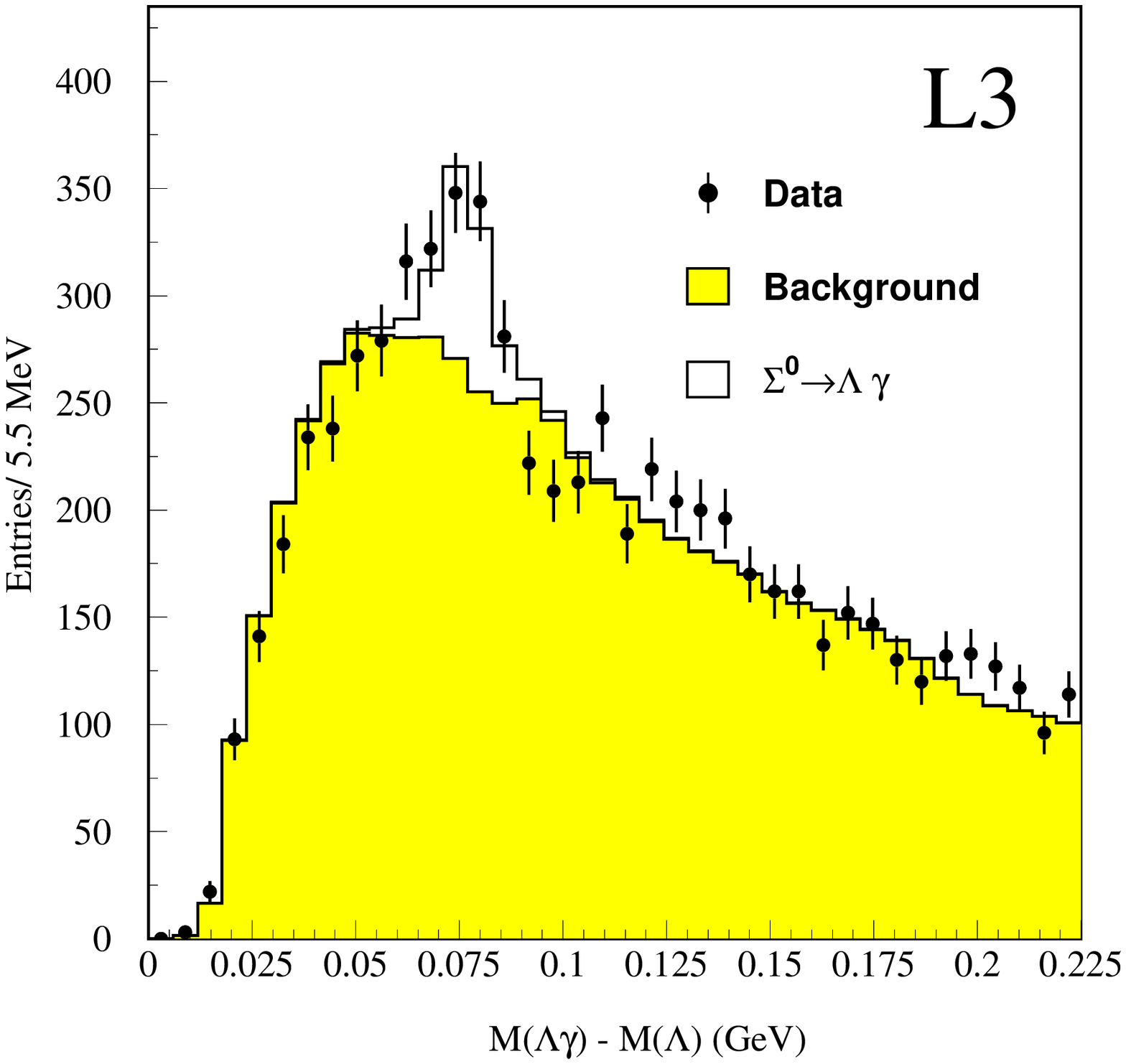}
\caption{The mass difference distribution
$\rm M(\Lambda\gamma) - M(\Lambda)$
of $\Sigma^0$ candidates.
The shapes of the signal and of the background are obtained from
Monte Carlo simulation as described in the text.}
\label{fig:newsz}
\end{center}
\end{figure}

Figure~\ref{fig:newsz} shows the mass difference distribution of the selected
$\Lambda \gamma$ candidates.
The number of selected
$\Sigma^0 \rightarrow \Lambda (\rightarrow {\rm p} \pi^-) \gamma$ decays
is found to be 263 $\pm$ 42.
%
%
\section*{Rate Measurements}

A total of 6.2~million hadronic Z decays were simulated with the
JETSET Monte Carlo to estimate detector effects.
The simulation accounts for the run conditions in luminosity-weighted periods
throughout the data-taking.

For signal and background fitting, further tuning of Monte Carlo
distributions is performed.
The signal shape predicted by the simulation is adjusted to fit the
observed peak position and resolution measured in the data.  The change in
the rate for a corresponding change of 0.5 in
the negative log likelihood in the fit of the background and signal
normalizations is taken as a systematic error.

To estimate the statistical error on the Monte Carlo determination of the
background shape, the signal is fit with a fixed normalization of the
background corresponding to two cases: a) increasing the background-level
by one standard deviation (given by the two parameter fit), and b) decreasing
the background-level by the same amount.  In the case of the $\Sigma^0$,
where the photon energy spectrum goes down to threshold, the observed
difference in the low-energy photon energy spectrum between Monte Carlo and
data is used to reweight the mass difference distribution.  The correlation
on the rate measurement between reweighting based on background versus
background plus signal is quoted as a systematic error.

In order to estimate the error on the kinematic extrapolation of the
observed rate to the full phase space of the $\Sigma^+$ and $\Sigma^0$
baryons produced in the fragmentation process, several Monte Carlo
generators are used to predict the
extrapolation~\cite{ARIADNE,JETSETuse,HERWIGuse}.
The variations in the fractions of decays inside the kinematic ranges of the
$\Sigma^+$ and $\Sigma^0$ selections are given in Table~\ref{tab:extrapolate}.
These are taken as systematic errors.
Varying the $\Sigma^+$ and $\Sigma^0$ selection cuts
over a wide range of cut values
has no observed systematic effect on the measured rates.

Using the JETSET Monte Carlo as a reference for the extrapolation,
the efficiencies
for $\Sigma^+$ and $\Sigma^0$ are found to be 0.19\% and 0.18\%, respectively.
The error on the efficiencies from Monte Carlo statistics is taken as a
systematic error.
The sources of systematic errors in the production rate measurements
are listed in Table~\ref{tab:systerror}.
The average numbers of $\Sigma^+$ and $\Sigma^0$ per hadronic
Z decay are measured to be:
  \begin{eqnarray*}
    \left< N_{\Sigma^+} \right> + \left< N_{\bar{\Sigma}^+} \right> & = & 0.114 \pm 0.011_{\mbox{\it \small stat}} \pm 0.009_{\mbox{\it \small syst}} \\
    \left< N_{\Sigma^0} \right> + \left< N_{\bar{\Sigma}^0} \right> & = & 0.095 \pm 0.015_{\mbox{\it \small stat}} \pm 0.013_{\mbox{\it \small syst}} \ \mbox{.}
  \end{eqnarray*}

Our measurements of the $\Sigma^+$ and $\Sigma^0$ production rates in
hadronic Z decays are compared to baryon production rates predicted by 
JETSET, HERWIG and ARIADNE models in Table~\ref{tab:models}~\cite{l3-38}.
In this table, the numbers obtained from the string-based and the hadron gas
models are also listed.  The predictions of all of these models underestimate
our measured rates.
Our measurements are consistent with other measurements performed at
LEP~\cite{opal,delphi,aleph}.
However, we observe rates which are somewhat higher.
This difference could arise from the low momentum part of the baryon
production spectra.
The use of low energy photons in the measurements of the $\Sigma^+$ and
$\Sigma^0$ production rates allows the kinematical ranges covered by the
L3 measurements to extend down to lower baryon momenta than those
measured by the other LEP detectors.

%
%
\section*{Acknowledgements}

We wish to express our gratitude to the CERN accelerator divisions for
the excellent performance of the LEP machine. 
We acknowledge the contributions of the engineers 
and technicians who have participated in the construction 
and maintenance of this experiment.  
%
%

%
%
\newpage
\typeout{   }     
\typeout{Using author list for paper 203 -?}
\typeout{$Modified: Fri Feb  4 11:17:42 2000 by clare $}
\typeout{!!!!  This should only be used with document option a4p!!!!}
\typeout{   }
%
%
%
%
%
%

\newcount\tutecount  \tutecount=0
\def\tutenum#1{\global\advance\tutecount by 1 \xdef#1{\the\tutecount}}
\def\tute#1{$^{#1}$}
\tutenum\aachen            
\tutenum\nikhef            
\tutenum\mich              
\tutenum\lapp              
\tutenum\basel             
\tutenum\lsu               
\tutenum\beijing           
\tutenum\berlin            
\tutenum\bologna           
\tutenum\tata              
\tutenum\ne                
\tutenum\bucharest         
\tutenum\budapest          
\tutenum\mit               
\tutenum\debrecen          
\tutenum\florence          
\tutenum\cern              
\tutenum\wl                
\tutenum\geneva            
\tutenum\hefei             
\tutenum\seft              
\tutenum\lausanne          
\tutenum\lecce             
\tutenum\lyon              
\tutenum\madrid            
\tutenum\milan             
\tutenum\moscow            
\tutenum\naples            
\tutenum\cyprus            
\tutenum\nymegen           
\tutenum\caltech           
\tutenum\perugia           
\tutenum\cmu               
\tutenum\prince            
\tutenum\rome              
\tutenum\peters            
\tutenum\potenza           
\tutenum\salerno           
\tutenum\ucsd              
\tutenum\santiago          
\tutenum\sofia             
\tutenum\korea             
\tutenum\alabama           
\tutenum\utrecht           
\tutenum\purdue            
\tutenum\psinst            
\tutenum\zeuthen           
\tutenum\eth               
\tutenum\hamburg           
\tutenum\taiwan            
\tutenum\tsinghua          
{
\parskip=0pt
\noindent
{\bf The L3 Collaboration:}
\ifx\selectfont\undefined
 \baselineskip=10.8pt
 \baselineskip\baselinestretch\baselineskip
 \normalbaselineskip\baselineskip
 \ixpt
\else
 \fontsize{9}{10.8pt}\selectfont
\fi
\medskip
\tolerance=10000
\hbadness=5000
\raggedright
\hsize=162truemm\hoffset=0mm
\def\r{\rlap,}
\noindent

M.Acciarri\r\tute\milan\
P.Achard\r\tute\geneva\ 
O.Adriani\r\tute{\florence}\ 
M.Aguilar-Benitez\r\tute\madrid\ 
J.Alcaraz\r\tute\madrid\ 
G.Alemanni\r\tute\lausanne\
J.Allaby\r\tute\cern\
A.Aloisio\r\tute\naples\ 
M.G.Alviggi\r\tute\naples\
G.Ambrosi\r\tute\geneva\
% ...
\newpage
\begin{list}{A}{\itemsep=0pt plus 0pt minus 0pt\parsep=0pt plus 0pt minus 0pt
                \topsep=0pt plus 0pt minus 0pt}
\item[\aachen]
 I. Physikalisches Institut, RWTH, D-52056 Aachen, FRG$^{\S}$\\
 III. Physikalisches Institut, RWTH, D-52056 Aachen, FRG$^{\S}$
\item[\nikhef] National Institute for High Energy Physics, NIKHEF, 
     and University of Amsterdam, NL-1009 DB Amsterdam, The Netherlands
\item[\mich] University of Michigan, Ann Arbor, MI 48109, USA
\item[\lapp] Laboratoire d'Annecy-le-Vieux de Physique des Particules, 
     LAPP,IN2P3-CNRS, BP 110, F-74941 Annecy-le-Vieux CEDEX, France
\item[\basel] Institute of Physics, University of Basel, CH-4056 Basel,
     Switzerland
\item[\lsu] Louisiana State University, Baton Rouge, LA 70803, USA
\item[\beijing] Institute of High Energy Physics, IHEP, 
  100039 Beijing, China$^{\triangle}$ 
\item[\berlin] Humboldt University, D-10099 Berlin, FRG$^{\S}$
\item[\bologna] University of Bologna and INFN-Sezione di Bologna, 
     I-40126 Bologna, Italy
\item[\tata] Tata Institute of Fundamental Research, Bombay 400 005, India
\item[\ne] Northeastern University, Boston, MA 02115, USA
\item[\bucharest] Institute of Atomic Physics and University of Bucharest,
     R-76900 Bucharest, Romania
\item[\budapest] Central Research Institute for Physics of the 
     Hungarian Academy of Sciences, H-1525 Budapest 114, Hungary$^{\ddag}$
\item[\mit] Massachusetts Institute of Technology, Cambridge, MA 02139, USA
\item[\debrecen] KLTE-ATOMKI, H-4010 Debrecen, Hungary$^\P$
\item[\florence] INFN Sezione di Firenze and University of Florence, 
     I-50125 Florence, Italy
\item[\cern] European Laboratory for Particle Physics, CERN, 
     CH-1211 Geneva 23, Switzerland
\item[\wl] World Laboratory, FBLJA  Project, CH-1211 Geneva 23, Switzerland
\item[\geneva] University of Geneva, CH-1211 Geneva 4, Switzerland
\item[\hefei] Chinese University of Science and Technology, USTC,
      Hefei, Anhui 230 029, China$^{\triangle}$
\item[\seft] SEFT, Research Institute for High Energy Physics, P.O. Box 9,
      SF-00014 Helsinki, Finland
\item[\lausanne] University of Lausanne, CH-1015 Lausanne, Switzerland
\item[\lecce] INFN-Sezione di Lecce and Universit\'a Degli Studi di Lecce,
     I-73100 Lecce, Italy
\item[\lyon] Institut de Physique Nucl\'eaire de Lyon, 
     IN2P3-CNRS,Universit\'e Claude Bernard, 
     F-69622 Villeurbanne, France
\item[\madrid] Centro de Investigaciones Energ{\'e}ticas, 
     Medioambientales y Tecnolog{\'\i}cas, CIEMAT, E-28040 Madrid,
     Spain${\flat}$ 
\item[\milan] INFN-Sezione di Milano, I-20133 Milan, Italy
\item[\moscow] Institute of Theoretical and Experimental Physics, ITEP, 
     Moscow, Russia
\item[\naples] INFN-Sezione di Napoli and University of Naples, 
     I-80125 Naples, Italy
\item[\cyprus] Department of Natural Sciences, University of Cyprus,
     Nicosia, Cyprus
\item[\nymegen] University of Nijmegen and NIKHEF, 
     NL-6525 ED Nijmegen, The Netherlands
\item[\caltech] California Institute of Technology, Pasadena, CA 91125, USA
\item[\perugia] INFN-Sezione di Perugia and Universit\'a Degli 
     Studi di Perugia, I-06100 Perugia, Italy   
\item[\cmu] Carnegie Mellon University, Pittsburgh, PA 15213, USA
\item[\prince] Princeton University, Princeton, NJ 08544, USA
\item[\rome] INFN-Sezione di Roma and University of Rome, ``La Sapienza",
     I-00185 Rome, Italy
\item[\peters] Nuclear Physics Institute, St. Petersburg, Russia
\item[\potenza] INFN-Sezione di Napoli and University of Potenza, 
     I-85100 Potenza, Italy
\item[\salerno] University and INFN, Salerno, I-84100 Salerno, Italy
\item[\ucsd] University of California, San Diego, CA 92093, USA
\item[\santiago] Dept. de Fisica de Particulas Elementales, Univ. de Santiago,
     E-15706 Santiago de Compostela, Spain
\item[\sofia] Bulgarian Academy of Sciences, Central Lab.~of 
     Mechatronics and Instrumentation, BU-1113 Sofia, Bulgaria
\item[\korea]  Laboratory of High Energy Physics, 
     Kyungpook National University, 702-701 Taegu, Republic of Korea
\item[\alabama] University of Alabama, Tuscaloosa, AL 35486, USA
\item[\utrecht] Utrecht University and NIKHEF, NL-3584 CB Utrecht, 
     The Netherlands
\item[\purdue] Purdue University, West Lafayette, IN 47907, USA
\item[\psinst] Paul Scherrer Institut, PSI, CH-5232 Villigen, Switzerland
\item[\zeuthen] DESY, D-15738 Zeuthen, 
     FRG
\item[\eth] Eidgen\"ossische Technische Hochschule, ETH Z\"urich,
     CH-8093 Z\"urich, Switzerland
\item[\hamburg] University of Hamburg, D-22761 Hamburg, FRG
\item[\taiwan] National Central University, Chung-Li, Taiwan, China
\item[\tsinghua] Department of Physics, National Tsing Hua University,
      Taiwan, China
\item[\S]  Supported by the German Bundesministerium 
        f\"ur Bildung, Wissenschaft, Forschung und Technologie
\item[\ddag] Supported by the Hungarian OTKA fund under contract
numbers T019181, F023259 and T024011.
\item[\P] Also supported by the Hungarian OTKA fund under contract
  numbers T22238 and T026178.
\item[$\flat$] Supported also by the Comisi\'on Interministerial de Ciencia y 
        Tecnolog{\'\i}a.
\item[$\sharp$] Also supported by CONICET and Universidad Nacional de La Plata,
        CC 67, 1900 La Plata, Argentina.
\item[$\diamondsuit$] Also supported by Panjab University, Chandigarh-160014, 
        India.
\item[$\triangle$] Supported by the National Natural Science
  Foundation of China.
\item[\dag] Deceased.
\end{list}
}
\vfill


\clearpage

%
%

\begin{table}[hbt]
\begin{center}
\begin{tabular}{|c|c|c|}
\hline
Monte Carlo & 
$f^{\it inside}_{\Sigma^+}$ &
$f^{\it inside}_{\Sigma^0}$ \\ \hline
ARIADNE~4.08 &  0.317 & 0.100 \\ \hline
JETSET~7.4   &  0.321 & 0.097 \\ \hline
HERWIG~5.9   &  0.311 & 0.091 \\ \hline
\hline
Maximum Spread & 3.1\% & 9.0\% \\ \hline
\end{tabular}
\end{center}
\caption{Monte Carlo predictions of 
 the fractions of decays inside the kinematic ranges of the
$\Sigma^+$ and $\Sigma^0$ selections.}
\label{tab:extrapolate}
\end{table}

\begin{table}[hbt]
\begin{center}
\begin{tabular}{|c|c|c|}
\hline
$\begin{array}{c}
\mbox{Source of} \\
\mbox{Systematic Error}
\end{array}$ &
$\sigma_{\left< N_{\Sigma^+} \right> + \left< N_{\bar{\Sigma}^+} \right>}^{\mbox{\it \small syst}}$ (\%) &
$\sigma_{\left< N_{\Sigma^0} \right> + \left< N_{\bar{\Sigma}^0} \right>}^{\mbox{\it \small syst}}$ (\%) \\ \hline
$\begin{array}{c}
\mbox{Signal Shape} \\
\mbox{Resolution}
\end{array}$ &           6.1      &      7.0 \\ \hline
$\begin{array}{c}
\mbox{Signal Shape} \\
\mbox{Peak Position}
\end{array}$ &           0.6      &      1.9 \\ \hline
$\begin{array}{c}
\mbox{Background Shape} \\
\mbox{Statistics}
\end{array}$ &           2.4      &      6.4 \\ \hline
$\begin{array}{c}
\mbox{Background} \\
\mbox{Reweighting Procedure}
\end{array}$ &           --       &      2.9 \\ \hline
$\begin{array}{c}
\mbox{Kinematic} \\
\mbox{Extrapolation}
\end{array}$ &           3.1      &       9.0 \\ \hline
$\begin{array}{c}
\mbox{Monte Carlo} \\
\mbox{Signal Efficiency}
\end{array}$ &           3.5      &      3.8 \\ \hline \hline
$\begin{array}{c}
\mbox{Total} \\
\mbox{Systematic Error}
\end{array}$ &           8.1      &      14.1 \\ \hline
\end{tabular}
\end{center}
\caption{Sources of systematic error in the production rate measurements.}
\label{tab:systerror}
\end{table}

\begin{table}[hbt]
\begin{center}
\begin{tabular}{|c|c|c|} \hline
Monte Carlo & 
\rule{0pt}{12pt}
$\left< N_{\Sigma^+} \right> + \left< N_{\bar{\Sigma}^+} \right>$ &
\rule{0pt}{12pt}
$\left< N_{\Sigma^0} \right> + \left< N_{\bar{\Sigma}^0} \right>$ \\ \hline
JETSET  & 0.0711 & 0.0691 \\
HERWIG  & 0.0978 & 0.0692 \\
ARIADNE & 0.0716 & 0.0685 \\
string-based model & 0.0740 & 0.0774 \\
hadron gas model & 0.0732 & 0.0766 \\ \hline
\end{tabular}
\end{center}
\caption{Model predictions of $\Sigma^+$ and $\Sigma^0$ production rates in
e$^+$e$^-$ annihilations at 91~\GeV{}.
The predictions of the JETSET, HERWIG and ARIADNE models have been
obtained with parameters tuned to L3 global event shape distributions
and the charged-particle multiplicity distribution.}
\label{tab:models}
\end{table}

%
%

\end{document}